\documentstyle[11pt,psfig]{article}
\newcommand {\be}{\begin{equation}}
\newcommand {\ee}{\end{equation}}

\begin{document}

\begin{center}
{\Large\bf SOC in a population model with global control}\\[1 cm]

{\large\bf Hans-Martin Br\"oker$^1$ and Peter Grassberger$^{1,2}$}\\[0.5cm]

$^1$ Physics Department, University of Wuppertal, D-42097 Wuppertal,
Germany\\
$^2$ HLRZ c/o Forschungszentrum J\"ulich, D-52425 J\"ulich,
Germany\\[0.5cm]
\bf{\today}
\end{center}

\begin{abstract}
We study a plant population model introduced recently by J. Wallinga 
[OIKOS {\bf 74}, 377 (1995)]. It is similar to the contact process (`simple 
epidemic', `directed percolation'), but instead of using an infection or 
recovery rate as control parameter, the population size is controlled 
directly and globally by removing excess plants. We show that the model is 
very closely related to directed percolation (DP). Anomalous scaling laws 
appear in the limit of large populations, small densities, and long times. 
These laws, associated critical exponents, and even some non-universal 
parameters, can be related to those 
of DP. As in invasion percolation and in other models where 
the r\^oles of control and order parameters are interchanged, the critical 
value $p_c$ of the wetting probability $p$ is obtained in the scaling limit 
as singular point in the distribution of infection rates.
We show that a mean field type approximation leads to a model studied 
by Y.\-C.\ Zhang {\it et al.} [J.\ Stat.\ Phys.\ {\bf 58}, 849 (1990)]. 
Finally, we verify the claim of Wallinga that family extinction in a 
marginally surviving population is governed by DP scaling laws, and 
speculate on applications to human mitochondrial DNA.
\end{abstract}
\newpage

\section{Introduction}

The main concepts in the description of critical phenomena are the 
control parameter (e.g. the temperature $T$ or a wetting probability $p$) 
which is typically an intensive quantity, and the order parameter 
(magnetization, density, ...) which is typically the density of an 
extensive quantity. As the control parameter passes through the critical 
point, correlation lengths and times diverge and the order parameter 
shows a singular (albeit continuous) behavior.

These concepts apply not only to equilibrium physics but also to 
systems intrinsically off equilibrium. Maybe the best known example 
of a critical phenomenon which has no static analogue is directed 
percolation (DP) with the preferred axis taken as time $t$. In this 
interpretation, DP describes an epidemic process without immunization 
or death, the so-called `contact process' or `simple epidemic' 
\cite{kinzel,review,footnote}. The infection rate can be chosen as control 
parameter in this interpretation, and the density of infected individuals 
is the order parameter. The critical point corresponds then 
to a transition from a regime where any finite epidemic dies out finally, 
to a regime where the epidemic can grow sufficiently fast that it lives 
forever in spite of statistical fluctuations. Exactly at the critical 
point the size of the epidemic, both in terms of the number $N$ of infected 
individuals and in terms of its spatial extent $R$, increases with time, 
while the density $\rho \sim N/R^d$ decreases to zero.

But, as pointed out by P. Bak {\it et al.} \cite{bak}, there exist also 
nonequilibrium models which show anomalous scaling related to diverging 
length scales without possessing any obvious control parameter. This 
phenomenon, called self-organized critical (SOC), has 
been vigorously studied in the last ten years. It appears in models 
for sand piles \cite{bak}, earthquakes \cite{olami}, biological evolution 
\cite{bak-snepp} and forest fires\cite{dross,broe-grass}, among others.
A common signature of those models is that they are slowly driven into 
an unstable regime. In the limit of zero driving speed and infinite 
system size, the system is forced into a critical state regardless 
of the initial conditions. Another model which shows the same behavior 
(and which can therefore be regarded as a forerunner of SOC) is 
invasion percolation \cite{invasion}. 

Actually, there are substantial conceptual differences between different 
models with SOC. While in some of them (like the sand pile and forest fire 
models) the slow driving appears to be `genuine' and hard to replace by 
anything else without destroying criticality, this is not so in other models. 
Prototypes of the 
latter are invasion percolation and other models with `extremal dynamics' 
\cite{bak-snepp,snepp}. In these models, each site on a lattice carries 
some `fitness' $f$, and evolution takes place only at the site with 
minimal $f$. Obviously, one could in these models replace the evolution 
by one where {\it all} sites evolve whose fitness is below some critical 
value $f^\star$. This $f^\star$ would then play the r\^ole of control 
parameter, and critical behavior is found when $f^\star = f_c$. The 
close connection between this non-self organized critical models and 
their SOC cousins is demonstrated by the fact that fitness distributions 
in SOC models with extremal dynamics show singularities precisely at $f=f_c$. 

Therefore, some SOC models are just characterized by an interchange 
between order and control parameters \cite{sornette,grazha}. Indeed, 
this is true {\it cum grano salis} also in other SOC models like the 
sand pile model. Also there the essential feature is that control is not 
exerted via an intensive parameter, but directly on a global flux. 
In equilibrium systems, a similar situation would prevail if we would 
switch from a (grand-) canonical ensemble to a microcanonical ensemble. In 
general this has no substantial effect on equilibrium phase transitions and 
critical behavior, although there exist exceptions \cite{thirring,antoni}. 
On the other hand, it is e.g. well known that replacing voltage control by 
current control can have dramatic effects in nonlinear (and non-equilibrium!) 
electric circuits.

In the present paper we shall discuss a model which can be understood as 
`control-switched' directed percolation. Instead of controlling the 
infection rate as in ordinary DP, we control the size of the epidemic.
Essentially (with some minor differences outlined below) our model was 
introduced first by Wallinga \cite{wall} as a model for a population of 
weeds. In this model a farmer's field is treated as a regular 
$d$-dimensional lattice of size $L^d$, with at most one weed per lattice 
site. Seeds from these plants fall onto all neighboring sites
and in spring a new weed is growing on all those sites. 
This alone would lead to supercritical growth of the population. But 
the farmer has decided that he will not tolerate more than $N$ weeds 
in the entire field. Thus he eradicates all plants in excess of this 
number. He does this in a completely random way, by picking plants from 
random sites until exactly $N$ plants are left over.

The critical regime in this model is reached for $N\to\infty$ and $L\to 
\infty$, with $\rho = N/L^d\to 0$. As we shall see in the next section, 
the case of finite $N$ and $\rho = 0$ corresponds to subcritical DP, while 
that of $N=\infty$ and finite $\rho$ is equivalent to supercritical DP.

The model will be defined more formally in the next section. There, we 
shall also formulate scaling laws, and test them in one spatial dimension 
(simulations were also performed for $d=2$, but will not be shown here). 
We will find excellent agreement, with one notable exception. This 
exception shows that we have not yet fully understood the model.
On the other hand, some of the observed scaling laws have also 
non-trivial consequences for ordinary DP. 
A discussion of our results and a comparison with previous results is
presented in sec.3. Among others, we will treat there a mean 
field type approximation where we relax the constraint that each site is 
occupied by one plant at most. We will see that this leads to a model 
studied by Zhang {\it et al.} \cite{zhang} in which all critical exponents 
assume their mean field values.

\section{The model and its scaling laws}

\subsection{The model}

Consider a population of $N$ plants distributed on a $d$-dimensional
hypercubic lattice of linear size $L$, such that any site is either empty 
or occupied by one plant. Boundary conditions are assumed to be periodic. 
The reproduction and distinction of the plants
happens in discrete time steps and is controlled by the following rules:

(0) At the beginning the $N$ plants are randomly distributed onto
the lattice.

(i) At each time step $t$ every plant places $2d$ seeds at its
neighboring sites and dies. 

(ii) On every site which has received at least one seed a new plant is 
growing. Notice that this implies that all seeds die which fall onto sites 
which contain already another seed. Nevertheless, it is easy to see that 
the number $M(t)$ of new plants is strictly larger than $N$.

(iii) The size of the population is kept constant by randomly removing 
$M(t)-N$ plants, and leaving $N$ in place. The fraction of surviving plants 
is denoted by ${\tilde p}(t)=N/M(t)$.

This model differs from that in \cite{wall} in the assumption that seeds 
fall only onto neighbors of the parent plant, not onto the site of the 
parent plant itself. This is done in order to make closer contact to 
existing treatments of DP. 

A variant of this model is obtained by replacing steps (i)-(iii) by the 
following steps:

(i') Plants are chosen at random, and for each chosen plant a random nearest
neighbor is chosen as well. If this neighbor had already been chosen in the 
same time step, another plant is chosen. This is repeated until a free 
neighbor is found, and a seed is put there. 

(ii') This is repeated until $N$ seeds have been placed.

(iii') After that, each of the seeds grows into a new plant. The number of 
bonds (= pairs of different neighbors) tested during this time step is denoted 
as $M(t)$, and ${\tilde p}(t)=N/M(t)$.

Just as the first variant will be seen to correspond to directed site percolation, 
this variant corresponds to bond percolation. 

In the following we shall present numerical results from simulations for 
$d=1$. We have also made simulations in two dimensions. They show essentially 
the same results. But they are less significant since deviations from mean field 
behavior are less pronounced, they are harder to analyze, and they are more 
affected by finite size corrections. Therefore we shall not present them 
in this paper. We shall present results only for the first variant, although 
again we have also performed extensive simulations for the second variant.

Actually, instead of starting with a random distribution, we started in some 
runs with all plants located at {\it even} sites of the lattice. As we shall 
show in subsec. 2.4, for $\rho <1/2$ and $t\to\infty$ only such configurations 
survive where the number of empty sites between two plants is odd. 
Starting already with such a configuration can then shorten transient times.

\subsection{Supercritical case}

Let us first study the case where $N$ is large, and $\rho=N/L^d$ is non-zero 
in the limit $N\to\infty$. In this case it is obviously irrelevant whether 
we control $N$ rigidly or just in average, with fluctuations of order 
$N^{1/2}$. Such a soft control can be implemented by killing new plants 
completely at random, with probability $0<1-p<1$ for each new plant. As 
long as $p$ is larger than the critical threshold for directed site percolation
(which is 0.705485 in $d=1$ \cite{jens} -- notice that $d=1$ in our notation 
corresponds to 2 dimensions of space-time), this will be exactly the case 
of supercritical DP. In this case correlations are indeed short range and 
fluctuations of $N$ are of order $N^{1/2}$, as we required. Thus we 
conclude that our model and supercritical DP coincide in the limit 
$N\to\infty,\; \rho={\rm const}>0$. 

In particular, in this limit the long time average of the fraction of 
surviving plants should coincide with the `wetting' probability. Let us 
denote the `equation of state' for DP as 
\be
   \rho = \Phi(p)
\ee
where $\rho$ is the density of infected sites in an infinitely extended 
epidemic. For $p\to p_c$ from above, one has
\be
   \Phi(p) \sim |p-p_c|^\beta \;.
\ee
We predict then that the time average of ${\tilde p}(t)$ satisfies exactly 
the same relations, 
\be
   \langle {\tilde p}\rangle \equiv \langle N/M(t) \rangle = \Phi^{-1}(\rho)
                     \label{super-scal}
\ee
and 
\be
   \langle {\tilde p}\rangle -p_c  \sim \rho^{1/\beta}
\ee
for $N\to\infty, \rho=N/L^d >0$. Deviations from eq.(\ref{super-scal}) for 
finite $N$ should be analytic, since the model is not critical. Indeed we 
found that these deviations decreased as $1/N$. The function $\Phi(\rho)$ 
obtained from eq.(\ref{super-scal}) agreed within statistical errors with 
that obtained from DP. A log-log plot showing $\rho$ against 
$\langle {\tilde p}\rangle -p_c$ for very large $N$ and $t$ is shown in fig.1
Also plotted in this figure is a line with slope $\beta = 0.2765$ \cite{jens}. 
We see perfect agreement.

\begin{figure}[ht]
\centerline{\psfig{file=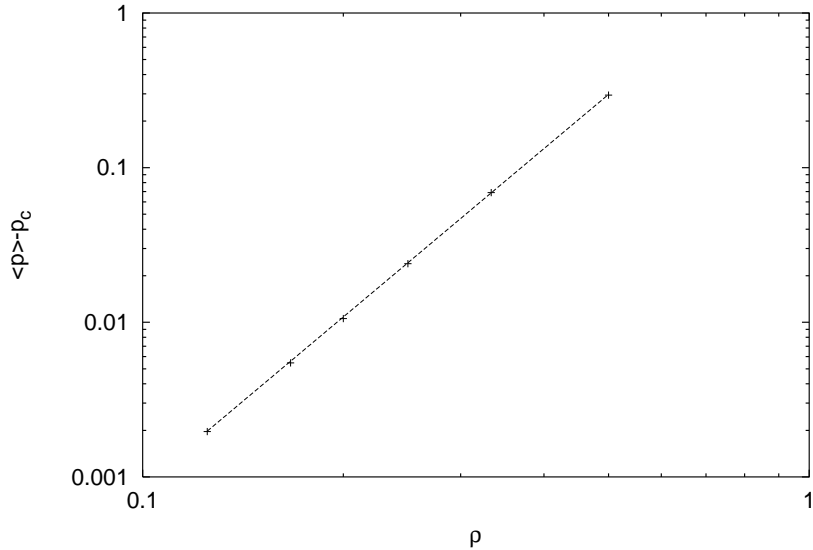,width=11cm,angle=270}}
\caption{
\small Double logarithmic plot of 
$\langle {\tilde p}\rangle -p_c $ versus $\rho$. The dashed line 
has the slope $1/\beta$ with $\beta = 0.2765$.
Here $N$ ranged from $1024$ $(\rho =1/8)$ to $ 2048 $ $(\rho =1/3)$.
Simulation times $t$ ranged from $10^8 $ $(\rho =1/8)$ to 
$5 \cdot 10^7 $ $(\rho =1/3)$, with the first $10^6$ iterations discarded.
Statistical errors are less than the sizes of the symbols.
} 
\label{rvpmpc.fig}
\end{figure}

In \cite{wall}, the author found $\langle {\tilde p}\rangle -p_c  \sim \rho^{1.66}$
for a model which should be in the universality class of DP in 2+1 dimensions. 
This is in reasonable agreement with the best estimate \cite{grazha} $\beta^{-1} 
= 1.71\pm 0.01$. We should, however, point out that we found for his model 
$p_c = 0.3178\pm 0.002$, while he had found $p_c= 0.36$. In view of this, the 
good agreement for the exponent is a bit surprising.

\subsection{Critical case}

The above argument tells us that the average survival rate tends 
to $p_c$ when $\rho\to 0$. In this limit fluctuations of $N$ become large in DP, 
and the very close relationship between DP and the present model breaks down.
Nevertheless, we shall argue that there is a non-trivial scaling law with 
an exponent determined by DP. 

\begin{figure}[ht]
\centerline{\psfig{file=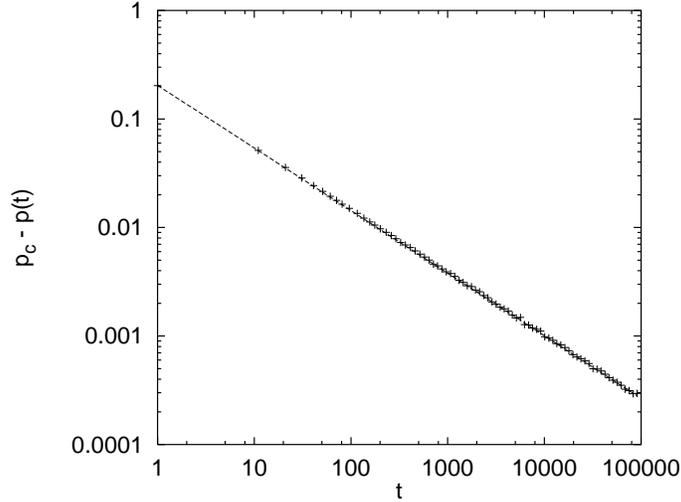,width=9cm,angle=270}}
\caption{
\small Double logarithmic plot of $p_c-{\tilde p}(t)$ versus $t$. Data 
are averaged over $100$ runs with $N= 2^{14}$ and $L= 100 \cdot 2^{14}$ 
each. The dashed line has slope $-1/\nu_{||}$ 
with $\nu_{||} = 1.7338$ \cite{jens}. 
} 
\label{pcmpvt.fig}
\end{figure}
 
Let us start at $t=0$ with a random, infinite, and infinitely dilute population 
of plants. Since in the first time step all plants have $2d$ free neighbors, 
the survival rate for the first step is ${\tilde p}(1) = 1/2d$. This is 
smaller than $p_c$ for any dimension. For the next time step it is no longer 
true that each plant has $2d$ free neighbors. Therefore $M(2)$ will be less 
than $2dN$, and ${\tilde p}(2) > 1/2d$. During successive time steps the 
clustering due to the locality of off-spring production will increase, and 
${\tilde p}(t)$ will increase with it, until ${\tilde p}$ reaches $p_c$ 
for $t\to\infty$. We expect thus a scaling law 
\be
   p_c - {\tilde p}(t) \sim t^{-\alpha}            \label{crit-scal}
\ee
with a yet unknown exponent $\alpha$. But $\alpha$ can be related to the 
exponent $\nu_{||}$ of DP if we assume that there is only one divergent time 
scale in critical DP, and that this time scale is uniquely related to $p-p_c$ 
as 
\be
   t \sim |p_c-p|^{-\nu_{||}}  \;.
\ee
If we identify here $p$ with ${\tilde p}$ in eq.(\ref{crit-scal}) then obviously 
\be
   \alpha = 1/\nu_{||} \;.       \label{alpnu}
\ee
Alternatively, assume we start with the same infinitely diluted initial 
population, but let it evolve according to DP. Since plants are infinitely 
far apart, their offsprings will evolve independently. If we would use $p=p_c$, 
the total population size would grow as $N(t) \sim t^\eta$ \cite{kinzel,review}. 
In order to prevent this and to keep at least $\langle $N(t)$ \rangle$ fixed, 
one has to use $t$-dependent values of $p$ which are slightly below $p_c$, 
$p=p_c - \epsilon(t)$ with $\epsilon(t)\to 0$ for $t\to \infty$. In order to 
obtain eq.(\ref{alpnu}), we again have to postulate that the relation between 
$\epsilon$ and $t$ is the same as between a fixed distance from the critical 
point and the corresponding correlation time.

Neither of these two arguments is of course rigorous, but simulations shown in 
fig.2 give again perfect agreement.

\subsection{Subcritical case}

The case $N<\infty$, $L\to \infty$ corresponds to the subcritical case of DP. 
In this case fluctuations of DP are strongest (indeed, if no precautions are
taken, all epidemics die with probability 1), and the connection between DP 
and the present model is the most delicate. It is indeed in this regime that 
we found problems that we have not yet fully understood.

For finite $N$ and $t\to\infty$, it is easy to see that all plants must have 
a common ancestor \cite{zhang}. Assume that there were $k$ different lines of ancestry, 
with $k>1$. Although the rules of the game prevent the total population from 
going extinct, they do not prevent any of the $k$ subpopulations from dying. 
Indeed, since their sizes will fluctuate, any of them has a finite chance 
to die, and thus $k$ decreases with a finite rate until $k=1$ is reached. 
Notice that the same argument leads also to the fact that plants survive 
only on one of the even/odd sublattices stated in sec.2.1.

\begin{figure}[ht]
\centerline{\psfig{file=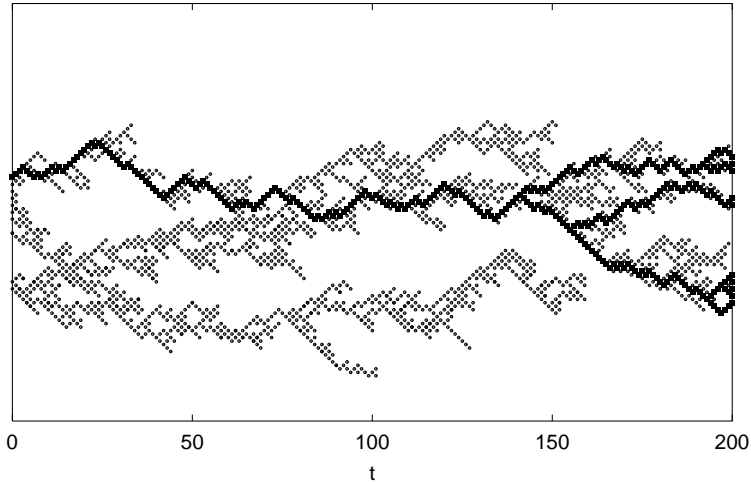,width=9.9cm,angle=270}}
\caption{
\small Typical run with $N=10$. Notice that all clusters die sooner or later 
except for one which is indicated by heavy dots. The surviving cluster
has an internal width $R$, a characteristic time $\tau$, and its
center of gravity follows essentially the random walk 
performed by the common ancestor. }
\label{conf.fig}
\end{figure}
 
Let us consider a finite population at some late time $t$. If we take any two 
plants $i$ and $j$, they will have a common ancestor $t_{ij}$ time steps back.
The maximum over $i$ and $j$ is the time it took the common ancester to wipe 
out all its competitors. 
This is a fluctuating number, its average value $\tau$ is assumed to scale 
with $N$, 
\be
   \tau = \langle \max_{i,j} t_{ij}\rangle \sim N^a      \label{a}
\ee
(see fig.3). Thus $\tau$ is a characteristic scale for the time it took for 
all other lines of ancestry to die out. 

During each time step, a daughter 
plant cannot move more than one lattice spacing from its parent. Therefore, 
all plants in the population must form a cluster of finite extent $R\leq \tau$
\cite{wall,zhang}. Again we assume a scaling law 
\be
   R \sim N^b \;,\qquad b\leq a\;.     \label{b}
\ee
Unless $b=1/d$, the cluster is fractal with dimension $d_f = 1/b$.

Notice that $\tau$ and $R$ must become stationary for large $t$. This is 
not so for the center of mass ${\bf X}$. Due to translation invariance, the 
difference $\Delta {\bf X}_T(t) = {\bf X}(t+T)-{\bf X}(t)$ is a stationary 
random variable for large $t$. Also, we should expect its autocorrelation 
time to be finite. Therefore, ${\bf X}(t)$ makes essentially a random 
walk \cite{zhang,torre}, 
\be
   \langle ({\bf X}(t+T)-{\bf X}(t))^2\rangle \sim D\;T \qquad {\rm for}\quad 
        T\to\infty    \;.       \label{walk}
\ee
For $N=1$ one easily sees that $D=1$. For large $N$ we again make a scaling
ansatz
\be
   D \sim N^c  \;.     \label{c}
\ee

Finally, we can look at the average value of ${\tilde p}(t)$, 
\be
   \langle{\tilde p}\rangle = \lim_{T\to\infty} {1\over T} \sum_{t=1}^T 
      {\tilde p}(t) \;.
\ee
For $N=1$ one obviously has $\langle{\tilde p}\rangle = 1/2d < p_c$, 
while ${\tilde p}\to p_c$ for $N\to\infty$. It seems thus natural to 
assume again a scaling law 
\be
   p_c - \langle {\tilde p} \rangle \sim N^{-x} \;.  \label{x}
\ee

Let us now try to predict these exponents in terms of DP critical exponents.
The closest analogy is with DP in the subcritical regime, but conditioned 
on those rare clusters which have not yet died out. Let us denote 
$\epsilon = p_c-p$. The general ansatzes for the average population size, 
survival probability, and spatial extent of DP clusters (or, rather contact 
model clusters; notice that in terms of DP proper we are actually dealing 
with cross sections of clusters at fixed $t$) are \cite{torre}
\be
   n(t,\epsilon) \sim \epsilon^{2\beta-d\nu_\perp} \psi(\epsilon^{\nu_\|} t) \;,
\ee
\be
   P(t,\epsilon) \sim \epsilon^\beta \phi(\epsilon^{\nu_\|} t) \;,
\ee
and
\be
   X(t,\epsilon) \sim \epsilon^{-\nu_\perp} \chi(\epsilon^{\nu_\|} t) \;.
\ee
Notice that we denoted the spatial extent by $X$, anticipating that a 
subcritical DP cluster has finite internal width, so that the total width 
is dominated by the spread of the center of gravity of the cluster.

For the long-time behavior at $\epsilon> 0$ one needs the behavior of the 
scaling functions when their argument $\epsilon^{\nu_\|} t \to \infty$. 
Assuming that surviving clusters perform random walks \cite{zhang,torre}, 
we have $X \sim \sqrt{t}$ and therefore
\be
   \chi(z) \sim \sqrt{z}\;, \qquad z\to\infty \;.
\ee
This gives us immediately 
\be
   D\sim \epsilon^{\nu_\| - 2\nu_\perp}    \label{Dp}
\ee
or 
\be
   {c\over x} = 2\nu_\perp - \nu_\| \;.   \label{cx}
\ee

Since we are interested only in the subcritical DP process conditioned 
to surviving clusters, we are not interested in $n(t,\epsilon)$ and 
$P(t,\epsilon)$ themselves, but only the ratio ${\bar N}(t,\epsilon) = 
n(t,\epsilon)/P(t,\epsilon)$. 
Since this has to converge to a finite constant for $t\to\infty$, we 
see that $\psi(z)/\phi(z) \to const$ for $z\to\infty$, and 
\be
   {\bar N} \sim \epsilon^{\beta-d\nu_\perp} \;,  \label{Np}
\ee
leading to the predictions
\be
   x = {1 \over d\nu_\perp - \beta}     \label{1overx}
\ee
and 
\be
    c = {2\nu_\perp - \nu_\| \over d\nu_\perp - \beta} ;.   \label{DN}
\ee

Equations (\ref{Dp}) and (\ref{Np}) could indeed have been obtained by 
simple dimensional arguments. In addition to assuming that there is only one 
characteristic time scale $\sim \epsilon^{-\nu_\|}$, we assume similarly 
that there is only one length scale $\sim \epsilon^{-\nu_\perp}$ and one 
characteristic density $\sim \epsilon^\beta$. Then, eq.(\ref{Dp}) 
follows simply by noting that $D$ has dimension [length]$^2$/[time], 
while eq.(\ref{Np}) follows from $[N] = [{\rm density}]\times[{\rm length}]^d$. 
Accepting this 
argument, we can immediately predict also the exponents in eqs.(\ref{a}) 
and (\ref{b}):
\be
   a = {\nu_\| \over d\nu_\perp - \beta} \;, \qquad b = {\nu_\perp 
               \over d\nu_\perp - \beta} \;.    \label{TRp}
\ee

\begin{figure}[ht]
\centerline{\psfig{file=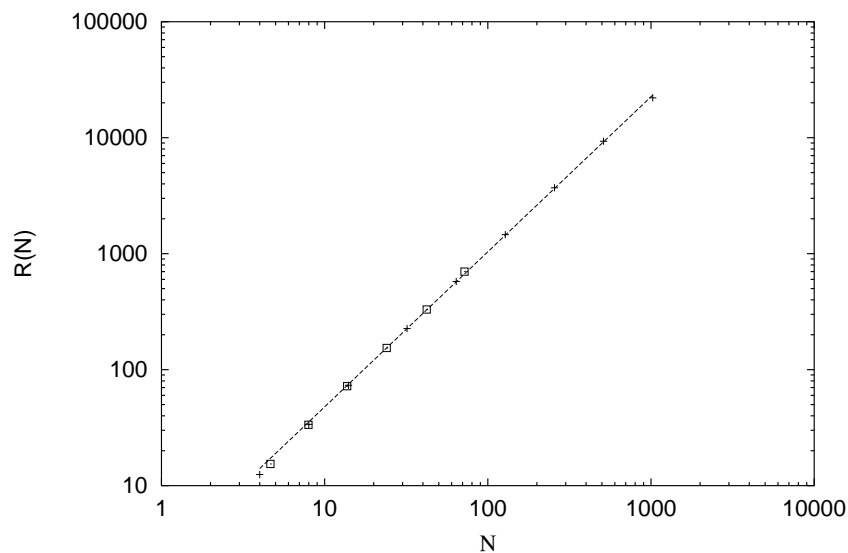,width=11cm,angle=270}}
\caption{
\small Double logarithmic plot of $R = \langle x_{\rm max} - x_{\rm min}
\rangle$ versus $N$ (crosses). In these simulations, 
long transients (typically $t = 10^6$) were discarded, and averages 
were taken over $10^8$ time steps. Statistical errors are smaller than 
the size of the symbols. The dashed line has slope $1.337$ as predicted by 
eq.(\ref{TRp}). We also show data for subcritical DP (rectangles). They 
not only increase with the same power of $N$, but within error bars they 
have even the same amplitude.
}
\label{r-n.fig}
\end{figure}
\vspace{.5cm}

\begin{figure}[ht]
\centerline{\psfig{file=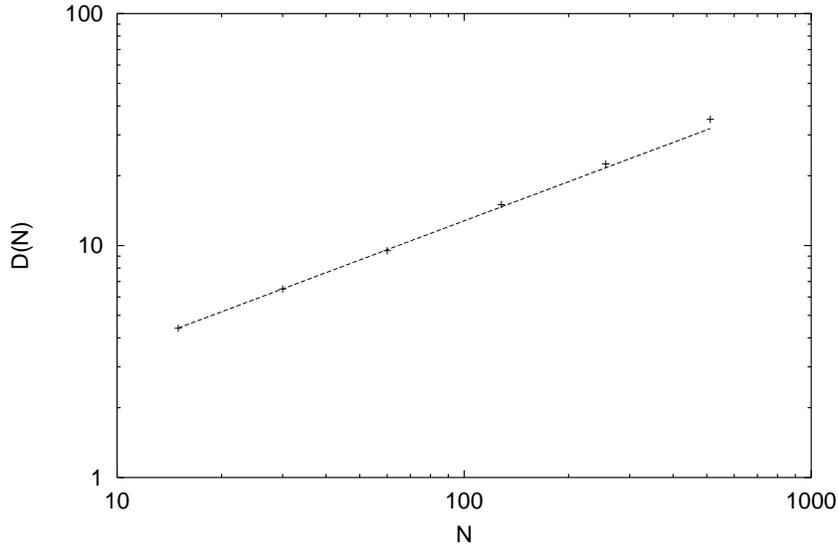,width=11cm,angle=270}}
\caption{
\small Double logarithmic plot of the diffusion coefficient $D$ versus 
$N$. The dashed line has slope $0.561$ as predicted by eq.(\ref{DN}). 
} 
\label{dvonn.fig}
\end{figure}

Again we show only numerical tests for $d=1$, although we made also some 
simulations for $d=2$. In more than one dimension of 
space, the most natural definition of $R$ is in terms of the gyration radius 
(rms. spread) of the population cluster. In $d=1$ an alternative is the 
end-to-end spread of the cluster, $R = \langle x_{\rm max} - x_{\rm min}
\rangle$. We 
checked that both definitions give compatible results. Results obtained 
with the latter definition are shown in fig.4. In this figure we also 
show data for subcritical DP, obtained with an enrichment trick
\cite{wall-erp}: in order to compensate the exponential decrease of the 
number of surviving clusters, we made copies of these clusters in regular
time intervals. By judiciously adjusting the copying rate, one can keep the 
number of surviving clusters roughly constant, allowing them to be 
studied up to very long times. Surprisingly, we found that the DP data 
not only showed the same scaling as the data of the present model (this 
was expected), but they have also the same amplitude within the measured 
error bars. We don't have any good explanation for this.

Results for the diffusion coefficient $D$ obtained from the same runs 
are shown in fig.5. Notice that $D$ increases with the population size. 
Naively, one might have guessed that it would decrease. The latter would 
be true if the wandering of the cluster were a cooperative effect. But as 
we have seen, it is due to the diffusive motion of a single plant line, 
namely the common ancestor of the entire cluster. This line of ancestry 
is singled out by having had better chances of survival than its competitors. 
Without local saturation (i.e. the fact that seeds falling onto sites which 
contain already a seed are lost), each line would have had the same 
chance of survival, and $D$ would be independent of $N$ \cite{zhang}. 
With local saturation it is clear that a plant in a less dense region 
has a larger chance to have long lasting offspring. Thus survivors 
will generally be seeds which fall further from the center of the 
cluster, and which lead thus to faster diffusion. 

The fact that $D$ increases with $N$ holds also for $d>1$. It seems at odd
with the observation that ``clusters of weeds tend to remain on the same 
spot for a long time" of \cite{wall}, but this shows just that diffusion is 
a slow process and can be easily missed if the diffusing object itself is 
fuzzy and changing its shape.

\begin{figure}[ht]
\centerline{\psfig{file=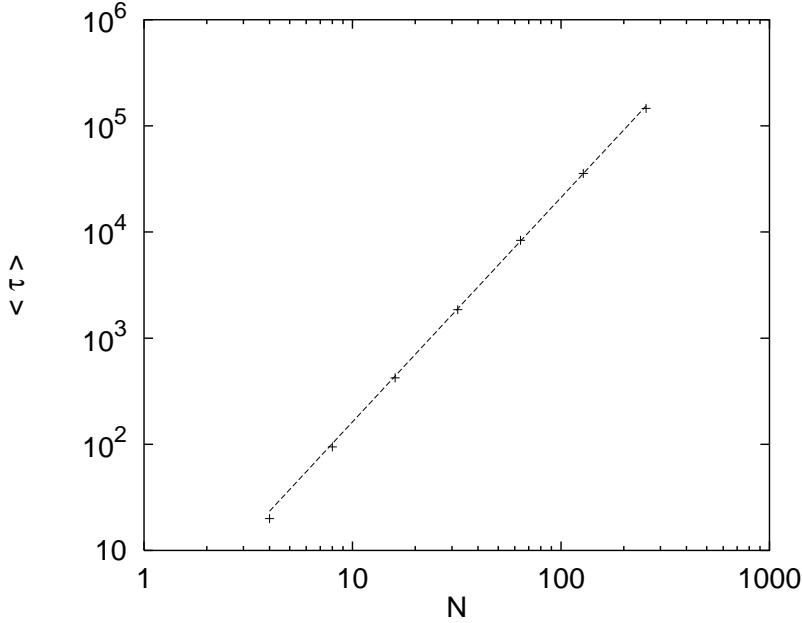,width=11cm,angle=270}}
\caption{
\small Double logarithmic plot of $\langle \tau\rangle$ versus
$N$. The dashed line has slope $2.114$ as predicted by eq.(\ref{TRp}).
Statistical errors are again smaller than the size of the symbols.
} 
\label{tvonn.fig}
\end{figure}

Instead of showing data for $\tau$ defined exactly as above, we show in fig.6 
data for a slightly different average. Instead of averaging over all times 
$t$, the data shown in fig.6 were obtained as follows. To each plant we 
attach a label which is inherited to all successors. We start with different 
labels for all plants and iterate until the entire population carries the 
same label. At this moment we measure $\tau$, relabel the plants so that 
each has again a different label, and continue with the evolution. Thus 
the results shown in fig.6 present averages taken only at those times when 
all plants have for the first time a common ancestor. While this should affect 
the prefactor in eq.(\ref{a}), it should not affect the exponent. We see 
perfect agreement with the prediction.

\begin{figure}[ht]
\centerline{\psfig{file=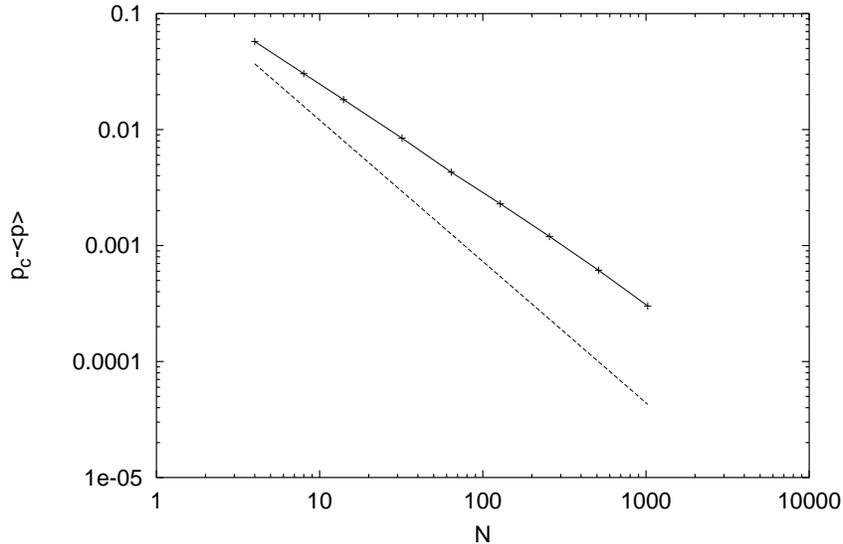,width=11cm,angle=270}}
\caption{
\small Double logarithmic plot of $p_c-\langle {\tilde p}\rangle $ versus
 $N$. The dashed line indicates the exponent predicted by eq.(\ref{1overx}).
 Obviously it does not fit the data.  
}
\label{pc-p(n).fig}
\end{figure}

\begin{figure}[ht]
\centerline{\psfig{file=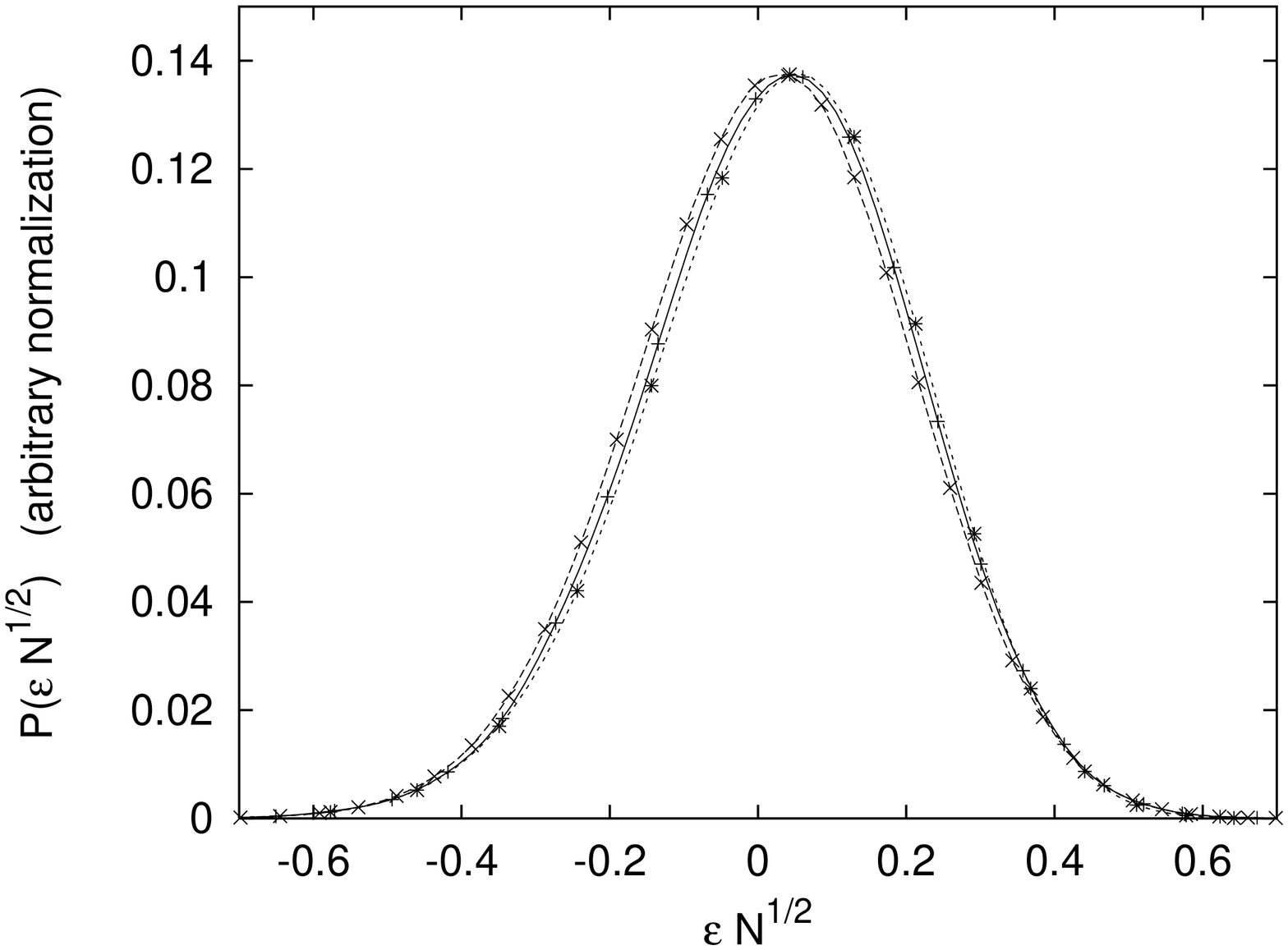,width=11cm,angle=270}}
\caption{
\small Plots of the distribution of ${\tilde p}(t)$ versus 
$(p_c-{\tilde p}(t))\sqrt{N}$, for $N=30$ ($\star$), $N=60$ ($+$), and $N=120$ 
($\times$). The lines are spline fits to the data 
points. Normalization is arbitrary.
}
\label{p-distrib.fig}
\end{figure}

On the other hand, it seems that eq.(\ref{1overx}) is wrong, at least if we 
identify $\epsilon$ with the difference $p_c-\langle {\tilde p}\rangle $. 
Simulations show a clear power law dependence of 
$p_c - \langle {\tilde p}\rangle$ on $N$ (see fig.7), 
but the exponent seriously disagrees with the prediction. Numerically we 
found $x=0.94 \pm 0.02$, while eq.(\ref{1overx}) would give $x = 1.2191$.
This suggests that simple linear averaging is not the 
right procedure to obtain the characteristic deviation of ${\tilde p}(t)$
from $p_c$ \cite{footnote2}. We checked that taking geometric or harmonic averages of 
${\tilde p}(t)$ did 
not improve the situation, since they gave essentially the same results. 
Rather, it seems that one should take some non-linear average of the 
difference $p_c - {\tilde p}(t)$. A priori, the most natural candidate 
might seem the geometric average of $p_c - {\tilde p}(t)$, but this is ruled 
out by the fact that $p_c - {\tilde p}(t)$ is not positive definite. 
Distributions of $p_c - {\tilde p}(t)$ are plotted in fig.8 for several 
values of $N$. We see that the average value is positive, but as $N$ 
increases the width of the distribution becomes much larger than this 
difference. In this limit, the distribution approaches a Gaussian 
whose width scales roughly as $1/\sqrt{N}$.

\subsection{Offspring Statistics, Family Survival, and Mitochondrial Eve}

Let us consider again the critical case, and let us label at some 
given time $t_0$ all plants with different labels. We can then study the 
evolution of the labeling pattern. As some branches of the population 
die, this pattern will coarsen increasingly. We are in particular 
interested in the limit where the coarsening is followed during a time 
$T$ which is $\gg 1$ but much smaller than $t_0$. In this limit the 
system can be considered as stationary during $t_0 < t < t_0+T$, although 
it of course never becomes strictly stationary at the critical point.

This is essentially the famous Galton - 
Watson problem of survival of family names \cite{branch-proc}, but in 
a marginally surviving total population with local competition and 
local off-spring production. This problem was studied in \cite{wall}, 
and it was mainly because of the results obtained thereby that this 
author claimed that his model is in the DP universality class.

Other phenomena and models related to this problem are the voter model 
\cite{liggett}, and the mitochondrial Eve problem \cite{eve,hase}. The voter 
model is essentially the extremely supercritical version where each lattice 
site is occupied by a plant, and sites receive in the next generation a 
seed from a randomly chosen neighbor. Its name derives from a
caricature of political opinion formation where a citizen adopts at
each time step the former opinion of one of his $2d$ neighbors. Its
upper critical dimension is $d=2$. For $d > 2$, the number of
different opinions in a large population with no two individuals
sharing originally the same opinion decrease as $N_{\rm opinion} \sim
1/t$. This is the same as in the Galton - Watson problem which is
essentially the mean field (or random neighbor) version of this
model. For $d=2$ one has logarithmic corrections, $N_{\rm opinion}
\sim t^{-1} \ln t$, and for $d=1$ one has $N_{\rm opinion}
\sim 1/{\sqrt{t}}$.  

The Eve problem arises from the 
fact that mitochondrial DNA (mtDNA) is inherited only from mother to child, 
without any contribution from the father. Thus different strands of 
mtDNA act like different labels in our model. Variations present humans 
suggest that all humans have a surprisingly recent common female
ancestor, called mitochondrial Eve, who lived ca. $200,000$ years
(i.\ e.\ $100,000$ generations) ago in southern Africa. The problem is whether 
this is simply explained as a statistical effect (for any single trait, 
one expects that only one line of ancestry survives finally), or whether 
this points to a common origin of the human race. This depends on
whether the characteristic time $\tau$ discussed in the last
subsection - which can be defined also for the voter and Galton -
Watson processes - is smaller than $10,000$ generations or not. 
We should mention that similar arguments apply also to the Y
chromosome which is inherited only from the 
father \cite{dorit}.

If the human populations would have been well mixed and of stationary
size during most of the last 100,000 generations, we could use the
Galton - Watson process to argue that a common root at about this time
was to be expected if the population size was $N < 100,000$. In this
case it would be a mere coincidence that this root is in Africa, and
ancestry lines involving males would most likely point to other
origins. If, however, the population size was larger, then this
finding is non-trivial and could be only explained by important
migrations. It would then point to a relatively recent African origin
of all humans. If we base our estimate on the voter model, on the
other hand, we would estimate $N\sim 1100$ as critical population
size, and the findings of~\cite{eve,hase} would seem even more
significant.

Let us now come back to the present plant model. The mitochondrial Eve
problem deals indeed with a finite population, treated in the last
subsection. But one can also study subpopulations with identical
traits (e.\ g.\, identical mtDNA) in an infinite population, and that
is what we shall do in the following.  
Since we are interested in this problem exactly at the critical point, 
the following discussion applies in exactly the same way to DP.

We denote by $P(t)$ the survival probability 
of a family, and $N(t)$ the average size of surviving families.  
Dealing with a critical phenomenon, we expect of course that both
should follow power laws. Due to stationarity,
the average number of plants carrying the same label (i.e., the number 
of offspring in surviving families) must increase such that 
this decrease of the survival probability is exactly compensated, 
$N(t)P(t)= const$. 
Finally, we expect also that the size $R(t)$ occupied by a family 
should scale. 

On the other hand, our previous argument that there are unique length, 
time, and density scales suggests that \cite{wall}
\be
   R(t) \sim t^{\nu_\perp/\nu_\|}     \label{Rt}
\ee
and
\be
   N(t) \sim t^{(d\nu_\perp-\beta)/\nu_\|} \;  \label{nt}
\ee
Thus the spatial growth of a family cluster in a critical population 
should scale in the same way as an isolated population, and $N(t)$ 
should grow as for an isolated population conditioned on survival. 
This is not trivial; the analogous is not true, e.g., for the survival 
probability $P$. While an isolated 1-d population dies with an exponent 
$\delta = -\beta/\nu_\|= 0.16$, a family has to compete with other families 
and dies with the same exponent as $1/N(t)$.

\begin{figure}[ht]
\centerline{\psfig{file=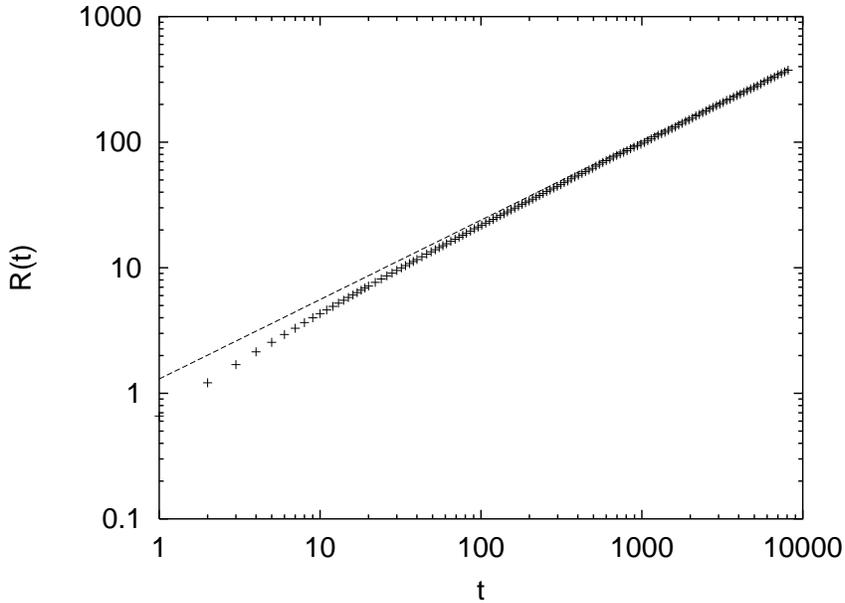,width=11cm,angle=270}}
\caption{
\small Log-log plot of $R(t)$ versus $t$, where $R(t)$ is the average end-to-end 
size of families in a stationary critical population. The dashed line has 
the slope 0.6326 predicted by eq.(\ref{Rt}). Error bars are again smaller than 
symbol sizes.}  
\label{Rt.fig}
\end{figure}

\begin{figure}[ht]
\centerline{\psfig{file=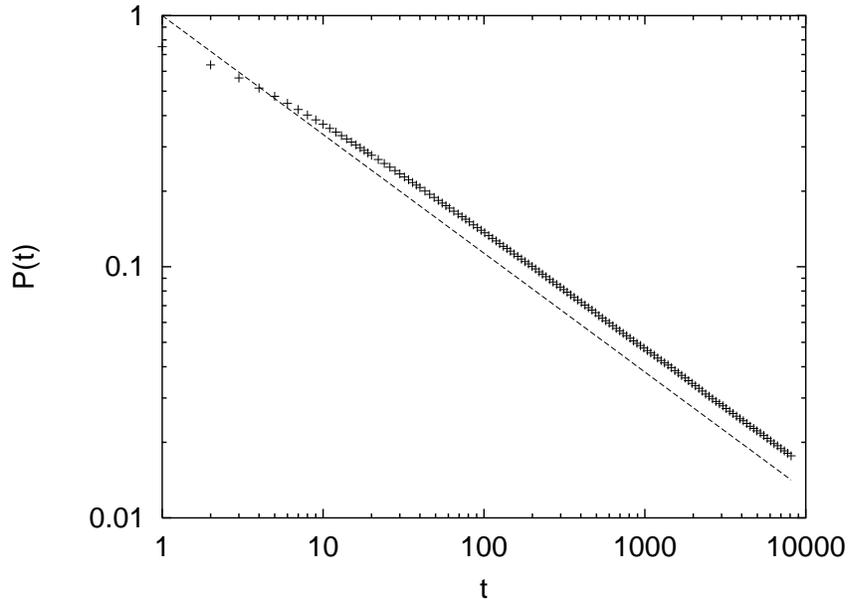,width=11cm,angle=270}}
\caption{
\small Log-log plot of $P(t)$ versus $t$, where $P(t)$ is the survival 
probability of a family. The dashed line has the slope $-0.4731$ predicted by 
eq.(\ref{nt}). The average family size increases as $1/P(t)$. Error bars are 
again smaller than symbol sizes.
}
\label{Pt.fig}
\end{figure}

To test these predictions, we show numerical results (again for $d=1$) in 
figs. 9 and 10. These data were obtained with $N=8192$ and $L=2^{23}$. 
After discarding a transient with $t_0= 5\times 10^4$, plants were labeled 
and followed over 8192 time steps. After that each plant was given again 
a different label, and the population was followed for another 8192 time
steps, etc., for a total of $3\times 10^5$ time steps.
The data shown in figs. 9 and 10 are in very 
good agreement with the prediction. Small discrepancies can be explained 
as finite size effects. Notice that these data seem at first sight to 
be inconsistent. For $d=1$ one can easily see that families must 
occupy non-interleaved territory. If two plants at sites $x_1$ and 
$x_2$ carry the same label, all plants with $x\in [x_1,x_2]$ must also 
carry the same label. Thus, $R(t)$ must be smaller than the average 
distance between families. The latter grows as $1/P(t) \sim t^{0.473}$, 
while $R(t)$ grows faster, $R(t)\sim t^{0.633}$. The answer to this 
seeming contradiction is that DP clusters are fractals. They contain 
holes of all sizes, and thus $R(t)P(t) \ll 1$ for small $t$.

Finally we want to mention that $(d\nu_{\perp}-\beta)/\nu_{\parallel}
=0.68$ for $d=2$\cite{grazha}. Thus, if we assume the present model
for the human population during the last 200,000 years, the critical
population size for the observed mitochondrial Eve phenomenon would be
$N\approx 500$. This estimate is of course highly uncertain, but it
indicates again that an estimate based on the Galton - Watson process
is most likely too high.

\section{Discussion}

We studied a population model which can be understood as a variant of 
the simple epidemic model without immunization. The latter is in the same 
universality class as DP in $d+1$ dimension. Numerically, we studied 
in most detail that version in one dimension of space which corresponds 
to directed site percolation on the square lattice. Although slightly 
different in details, this model was first introduced as a model for 
weeds \cite{wall}. The crucial point is that it is not the fertility of 
the weeds which is controlled, but control is exercised directly on the 
population size. 

In contrast to DP one does not, therefore, control the `wetting'
probability. Instead, the global population size (i.e., the integral 
over the order parameter) is kept fixed, and the usual control 
parameter is allowed to 
fluctuate. This is analogous to controlling a thermal phase transition 
not by means of the temperature but by means of the energy. While such 
an exchange of control makes usually no big difference in equilibrium 
critical phenomena, it leads to SOC in the present case. Even without 
fine tuning any parameter, critical behavior is found in the limit 
of large populations and low densities.

Although our model is not governed by extremal dynamics, it shares 
with such models (as, e.g., invasion percolation) the fact that the 
threshold of the standard (non-self-organized) version is recovered 
in the critical limit. 

The same features are also shared with a recent model studied by
Tadic and Dhar \cite{tadic}. In both models the system is driven 
into the critical state of DP without any fine tuning. In both, the 
dynamics is not extremal. Finally, in both models some scaling 
laws and critical exponents are identical to those of DP, while 
others are only related to DP. But the detailed mechanisms which 
drive the system into the critical state are completely different in 
both models.

A more direct relation exists to a model studied by Zhang {\it et al.}
\cite{zhang}. The main difference between that model and ours is that 
these authors have assumed no local saturation. Any seed has the 
same chance to grow, whether it is far from other seeds or whether it 
fell into a densely crowded region. Thus population growth is {\it only} 
controlled globally in the model of Zhang {\it et al.}, while there is 
a local aspect in our model by the requirement that each lattice site 
is occupied by at most one plant. Thus the model of \cite{zhang} can 
be considered as a mean field version of ours where the average offspring 
number is the same for all plants, independent of their location with 
respect to the bulk of the population. It is easily seen that all scaling 
laws discussed in the present paper are satisfied in the model 
of Zhang {\it et al.}, but with `mean field' exponents $\nu_\| = 1, \;
\nu_\perp=1/2$, and $d\nu_\perp-\beta = 1$ \cite{footnote3}.
Thus, the diffusion of the cluster is independent of its size, and 
$R \sim \sqrt{N},\; \tau \sim N$. On the other hand, $p$ is 
independent of $N$ in the model of \cite{zhang}. This shows again that
the relation between  $p_c - \langle {\tilde p}\rangle$ and $N$ is 
not as robust as the other scaling relations.

We verified most of the results of \cite{wall}, but formulated and 
tested the scaling laws in much more details. The main true discrepancy 
is the fact that the diffusion of the population cluster in the 
case of finite populations was missed completely in \cite{wall}. 
On the other hand we verified the non-trivial claim of \cite{wall} 
that the growth of surviving labeled clusters inside a large 
critical population obeys exactly the same scaling laws as the growth 
of surviving isolated clusters.
in 2 dimensions of space this means that specific traits die out 
in a marginally surviving population with probability $\sim 1/t^{0.68}$. 
This is to be compared to $\sim 1/t$ for the Galton-Watson process, 
and to $\log(t) /t$
for the voter model \cite{liggett}, i.e. for the case of a 
supercritical DP. 

Studies of human mitochondrial DNA indicate a common female African
ancestor some $200,000$ years ago\cite{eve,hase}. The last result
suggests that this might be more significant in favor of a recent
African origin than one might have argued on the basis of the Galton -
Watson process.

\bigskip
\bigskip
\bigskip

Acknowledgement:\\
We thank W.\ Nadler for very useful discussions.
This work was supported by the DFG within the ``Graduiertenkolleg
Feldtheoretische und numerische Methoden in der Elementarteilchen-
und Statistischen Physik``, and within Sonderforschungsbereich 237.

\eject


\begin{thebibliography}{99}

\bibitem{kinzel} W. Kinzel: in {\it Percolation Structures and Processes}, 
   ed. by G. Deutscher, R. Zallen, and J. Adler (Adam Hilger, Bristol 1983).

\bibitem{review} P. Grassberger, ``Directed Percolation: Results and Open 
   Problems", in {\it Nonlinearities in Complex Systems}, ed. by S. Puri 
   and S. Dattagupta (Narosa, New Delhi 1997).

\bibitem{footnote} Actually, the contact process is defined for continuous time, 
   while DP and the model studied in the present paper have discrete time. Since 
   we are interested only in critical behavior, we here neglect this difference.

\bibitem{bak} P. Bak, C. Tang and K. Wiesenfeld, Phys. Rev. A {\bf 38}, 364 (1988).

\bibitem{olami} Z. Olami, H.J.S. Feder and K. Christensen, 
   Phys. Rev. Lett. {\bf 68}, 1244 (1992).
 
\bibitem{bak-snepp} P. Bak and K. Sneppen, Phys. Rev. Lett. {\bf 71}, 4083 (1993).

\bibitem{dross} B. Drossel and F. Schwabl, Phys. Rev. Lett. {\bf 69}, 1629 (1992).

\bibitem{broe-grass} H.-M. Br\"oker and P. Grassberger, 
   Phys. Rev. {\bf E 56}, 4918R (1997).

\bibitem{invasion} D. Wilkinson and J. Willemsen, J. Phys. {\bf A 16}, 3365 (1983).

\bibitem{snepp} K. Sneppen, Phys. Rev. Lett. {\bf 69}, 3539 (1992).

\bibitem{sornette} D. Sornette, A. Johansen, and I. Dornic, J. Phys. (France) I
   {\bf 5}, 325 (1995).

\bibitem{grazha} P. Grassberger and Y.-C. Zhang, Physica A {\bf 224}, 169 (1996).

\bibitem{thirring} P. Hertel and W. Thirring, Ann. of Physics {\bf 63}, 520 (1970).

\bibitem{antoni} M. Antoni and A. Torcini, preprint cond-mat/9801008 (1998).

\bibitem{wall} J. Wallinga, OIKOS {\bf 74}, 377 (1995).

\bibitem{zhang} Y.-C. Zhang {\it et al.}, J. Stat. Phys. {\bf 58}, 849 (1990). 

\bibitem{jens} I. Jensen, J. Phys.\ A {\bf 29}, 7013 (1996). 

\bibitem{torre} P. Grassberger and A. de la Torre, Ann. of Physics {\bf } (1979).

\bibitem{wall-erp} F.T. Wall and J.J. Erpenbeck, J. Chem. Phys. {\bf 30}, 634, 637 (1959).

\bibitem{footnote2} Notice that averaging was not essential in the supercritical and 
   critical cases, since there $N\to\infty$. If averaging was done there at all, it 
   was done only in order to reduce statistical errors.

\bibitem{branch-proc} K.B. Athreya and P.E. Ney, {\it Branching Processes} 
    (Springer, Berlin, 1972).

\bibitem{liggett} T.M. Liggett, {\it Interacting Particle Systems} (Springer, New York, 1985).

\bibitem{eve} L. Vigilant {\it et al.}, Science {\bf 253}, 1503 (1991).

\bibitem{hase} M. Hasegawa and S. Horai, J. Mol. Evol. {\bf 16}, 111 (1991).

\bibitem{dorit} R.L. Dorit, H. Akashi, and W. Gilbert, Science {\bf 268}, 1183 (1995).

\bibitem{tadic} B. Tadic and D. Dhar, preprint cond-mat/9707151.

\bibitem{footnote3} Notice that the latter does not agree with the standard mean 
    field value of $\beta$ in DP.


\end{thebibliography}
\end{document}